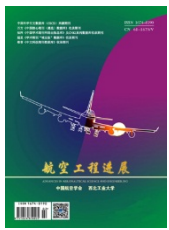

航空工程进展

*Advances in Aeronautical Science and Engineering*

ISSN 1674-8190,CN 61-1479/V
# 《航空工程进展》网络首发论文

| | |
|---|---|
| 题目： | 大型商用飞机单一飞行员驾驶的人因工程研究进展与展望 |
| 作者： | 许为，陈勇，董文俊，董大勇，葛列众 |
| 收稿日期： | 2021-03-13 |
| 网络首发日期： | 2021-09-10 |
| 引用格式： | 许为，陈勇，董文俊，董大勇，葛列众．大型商用飞机单一飞行员驾驶的人因工程研究进展与展望[J/OL]．航空工程进展．https://kns.cnki.net/kcms/detail/61.1479.V.20210910.1055.004.html |

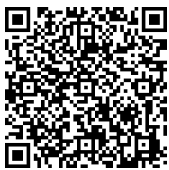

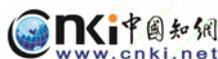
**网络首发**：在编辑部工作流程中，稿件从录用到出版要经历录用定稿、排版定稿、整期汇编定稿等阶段。录用定稿指内容已经确定，且通过同行评议、主编终审同意刊用的稿件。排版定稿指录用定稿按照期刊特定版式（包括网络呈现版式）排版后的稿件，可暂不确定出版年、卷、期和页码。整期汇编定稿指出版年、卷、期、页码均已确定的印刷或数字出版的整期汇编稿件。录用定稿网络首发稿件内容必须符合《出版管理条例》和《期刊出版管理规定》的有关规定；学术研究成果具有创新性、科学性和先进性，符合编辑部对刊文的录用要求，不存在学术不端行为及其他侵权行为；稿件内容应基本符合国家有关书刊编辑、出版的技术标准，正确使用和统一规范语言文字、符号、数字、外文字母、法定计量单位及地图标注等。为确保录用定稿网络首发的严肃性，录用定稿一经发布，不得修改论文题目、作者、机构名称和学术内容，只可基于编辑规范进行少量文字的修改。

**出版确认**：纸质期刊编辑部通过与《中国学术期刊（光盘版）》电子杂志社有限公司签约，在《中国学术期刊（网络版）》出版传播平台上创办与纸质期刊内容一致的网络版，以单篇或整期出版形式，在印刷出版之前刊发论文的录用定稿、排版定稿、整期汇编定稿。因为《中国学术期刊（网络版）》是国家新闻出版广电总局批准的网络连续型出版物（ISSN 2096-4188，CN 11-6037/Z），所以签约期刊的网络版上网络首发论文视为正式出版。

# 大型商用飞机单一飞行员驾驶的人因工程研究进展与展望

许为[1],陈勇[2],董文俊[2],董大勇[2],葛列众[1]

(1. 浙江大学 心理科学研究中心,杭州 310027)

(2. 中国商用飞机有限责任公司 上海飞机设计研究院,上海 201210)

**摘　要**：国内外民航界正在积极探索和研发大型商用飞机的单一飞行员驾驶(SPO)模式。针对SPO的人因工程研究也已初步展开,研究主要集中在驾驶舱机载设备升级方案、地面站飞行支持方案、"驾驶舱机载设备升级＋地面站飞行支持"的SPO组合方案。初步的人因工程研究倾向于SPO组合方案,但是,目前的人因工程研究尚不完善,无法为SPO提供完整的人因工程解决方案。本文综述了目前针对SPO的人因工程研究进展,分析了SPO人因工程的一些关键问题,指出目前研究中存在的问题和今后研究的重点,并针对今后的SPO人因工程研究提出了总体思路和建议。

**关键词**：单一飞行员驾驶;人因工程;以人为中心设计;民用飞机;飞机驾驶舱;人为因素

**中图分类号**：V323.11　　　　　　　　　　　　　　　　**文献标识码**：A

**开放科学(资源服务)标识码(OSID)**：

## Status and Prospect of Human Factors Engineering Research on Single Pilot Operations for Large Commercial Aircraft

XU Wei[1], CHEN Yong[2], DONG Wenjun[2], DONG Dayong[2], GE Liezhong[1]

(1. Center for Psychological Sciences, Zhejiang University, Hangzhou 310027, China;)

(2. Shanghai Aircraft Design and Research Institute, Commercial Aircraft Corporation of China Ltd., Shanghai 201210, China)

**Abstract**：The civil aviation community is actively exploring and developing the solutions of single pilot operations (SPO) for large commercial aircraft. Human factors engineering research for SPO has been launched, and the research mainly focuses on three research solutions: flight deck airborne equipment upgrade, flight support from ground stations, and the combined SPO solution of "flight deck airborne equipment upgrade ＋ flight support from ground stations". This paper reviews and analyzes the progress of human factors engineering research on SPO. The preliminary research outcome tends to support the combined SPO solution. However, the current human factors engineering research is not comprehensive and cannot provide a complete human factors engineering solution for SPO. For future human factors engineering research, this paper analyzes the key human factors issues on SPO and points out the gaps and the key areas for future work. Finally, this paper puts forward an overall strategy and recommendations for future human factors engineering research on SPO.

**Key words**：single pilot operations; human factors engineering; human-centered design; civil aircraft; flight deck; human factors

# 0 引 言

在过去50多年中,技术发展推动了大型商用飞机驾驶舱机组人员逐步递减(de-crewing)的趋势。从最初5名机组人员到目前机长和副驾驶2人配置,这种递减情况在不久的将来还会持续下去。在此背景下,作为新一代商用飞机发展核心技术之一,目前国内外民航界正在积极探索和研发大型商用飞机"单一飞行员驾驶"(Single Pilot Operations,以下简称SPO)模式。SPO指在大型民用飞机驾驶舱中仅配置一名飞行员(机长),借助提升的机载设备或者远程地面站操作员的支持(或者两者组合),能够在各种飞行场景中安全有效地完成航线飞行任务,并且达到不低于目前双乘员驾驶模式的飞行安全水平[1]。

支持SPO的人们认为SPO会导致一场航空运输革命,在满足当前商用飞机双乘员驾驶模式功能和安全性条件下,SPO可以带来减少飞行员数量、提升经济性、减少驾驶舱资源配置、缩小驾驶舱空间和减轻飞机重量等方面的好处。例如,42%的美国航线飞行员10年后将退休,累积的飞行员短缺在2035年将达到4万人[2]。SPO的这些优势对短途支线客货飞机表现得更加突出。然而,国际民航飞行员协会(ALPA)在2019年发布的白皮书《单人飞行操作的危险》中明确反对SPO,强调为维持飞行安全,大型民用飞机驾驶舱至少需要两名飞行员[3]。N. Stewart、D. Harris针对SPO的问卷调查表明,公众支持SPO的前提是保证飞行员健康、维持技术可靠性以及提供较大幅度的机票降价[4]。

美国NASA一直在持续开展针对SPO的相关研究,并且系统地提出了一些SPO总体设计方案[1],欧盟的"减少应激和工作负荷的高级驾驶舱"(ACROSS)计划也包括了SPO研究[5],Boeing,Airbus等飞机制造商以及Rockwell等航空设备供应商也在开展相关的SPO研发工作[3,6-8]。国内,针对SPO技术方案和系统架构的一些研究工作也开始起步[9-11],例如,2020年,上海交通大学王淼等[9]提出了一个SPO模式系统架构[9]。目前,国内一些航空公司也开始关注SPO[12]。

人因工程((Human Factors Engineering,民航界也称之为"人为因素")预备研究在飞机型号研发中起着无可替代的重要作用[13-14],SPO研发也不例外。作为一个标志性事件,2012年NASA的首次SPO技术交流会的主要议题之一就是关于SPO人因工程预备研究的重点和范围[1]。自那以后,在美国以NASA为主开展了一系列SPO人因工程研究[15-23],欧洲和澳洲等地的科研院校也开展了SPO人因工程研究[4,6,24-30],目前公开发表的SPO人因工程研究文献有数十篇,国内这方面的工作尚未启动。

人因工程界强调实现SPO的最大障碍不是技术本身,而是如何遵循"以人为中心设计"的理念,合理利用技术,研发出一个有效支持SPO系统设计和飞行安全的人因工程解决方案[13-14,24]。

针对SPO人因工程预研的重要性和迫切性,本文从人因工程角度出发,通过文献综述和分析来回答以下几个问题:目前国外的SPO人因工程预研进展到一个什么程度?有哪些主要结果?存在的主要问题是什么?在回答这些问题的基础上,本文深入分析了针对SPO的一些关键人因问题,提出了今后的研究重点。最后,针对今后SPO人因工程研究,本文提出了总的思路和建议。

# 1 SPO人因工程研究概述与现状

根据对现有公开发表的SPO人因工程研究文献的综述和分析,本部分首先概括目前SPO人因工程研究现状,然后在第二部分针对SPO关键人因问题逐一展开详细分析,找出目前研究中存在的问题,并且对今后的研究提出具体的建议。

## 1.1 SPO总体设计方案

开展SPO人因工程研究,首先要制定初步的SPO总体设计方案,然后通过一系列人因工程分析、建模、原型设计、实验研究等手段比较和验证所提出的SPO设计方案,确定最佳方案,最后为SPO系统设计提供人因工程的解决方案[1][15][24]。

初期的SPO研究主要涉及两种SPO总体方案[15-17][27]。一种是驾驶舱机载设备更新方案(以下简称驾驶舱方案),这是一种"以飞机为中心"的方案,主要通过提升现有驾驶舱机载自动化系统或引进机载智能自主化系统来替代现有人类副驾驶的部份职责,SPO飞行操作基本上不依赖于地面支持。

另一种是远程地面站支持方案(以下简称地

面站方案），这是一种"以空地为中心"的方案，具有分布式机组的设计概念[2]，将现有副驾驶的部份职责从空中移到了地面。在地面站方案中，地面站操作员主要承担以下三种角色：(1)常规的多机"签派员"；(2)为多架正常起飞或近进飞机提供支持的"港口飞行员"(harbor pilot)；(3)为处于非正常状态飞行(off-normal)的单架SPO飞机提供飞行支持的"远程飞行员"。因此，地面站操作员根据任务可分为：混合式操作员（承担以上所有三种角色）、专职地面操作员（承担角色1和2）、地面飞行员（承担角色3）。地面站方案也包括对驾驶舱机载设备以及地面站设备的技术升级。

随着研究的展开，研究者开始考虑第三种方案："驾驶舱＋地面站"组合式SPO总体设计方案。例如，NASA团队根据飞行员能力和飞行条件两个二维将该SPO组合式总体方案分成四个类别[15-16]（表1）。表1中的飞行条件是指除飞行员能力以外的影响飞行的因素，比如飞机、天气、机场等状态。驾驶舱机载和地面站设备更新可以是升级现有自动化系统或者引进新的智能自主化系统。

表1　SPO研究方案分类
Table 1　Taxonomy of SPO research solutions

| 飞行条件 | 飞行员能力 | |
| --- | --- | --- |
| | 正常飞行操作 | 非正常飞行操作 |
| 飞行员健康 | 类别1<br>· SPO驾驶舱机长操控飞行，驾驶舱机载系统辅助<br>· 地面站支持人员在地面站系统辅助下监控并支持多架SPO飞机 | 类别2<br>· SPO驾驶舱机长操控飞行，驾驶舱机载系统辅助<br>· 地面站支持人员在地面站系统辅助下为SPO飞机提供1对1"地面副驾驶"的遥控式飞行支持 |
| 飞行员失能 | 类别3<br>· 在地面站系统辅助下，地面站支持人员承担SPO驾驶舱机长职责，全权操控SPO驾驶舱机载系统，操控飞机安全着陆<br>· 驾驶舱机载系统执行来自地面站支持人员的命令 | 类别4<br>· 在地面站系统辅助和多名支持人员的辅助下，一名地面支持人员承担SPO驾驶舱机长职责，操控飞机安全着陆（除非飞机失去联系）<br>· 驾驶舱机载系统执行来自地面站支持人员的命令 |

如表1所示，从类别1到4，SPO的飞行操作变得更具挑战性，对实施SPO的飞行安全要求也相应提高，对SPO机长和地面站操作员的职责、设备支持、操作程序等方面的考虑也更为复杂。例如，类别1的SPO方案不需要大量的地面站支持。在类别2的的SPO方案中，SPO机长可能会要求地面站支持，尤其在非正常飞行场景以及工作量很大的情况下。类别3的SPO方案要求地面站操作员履行机长的职责，在驾驶舱机载设备系统的辅助下远程操控飞机安全着陆。在类别4的SPO方案中，履行机长职责的地面站操作员可能需要地面站其他人员的协助才能安全地操控飞机着陆。根据地面站操作员的角色，"签派员"需要支持方案类别1、2、3、4中的部分任务；"港口飞行员"支持类别1；"地面飞行员"需要支持类别2、3、4。

图1~图3分别演示了Airbus称之为"最小可行模拟器"的SPO飞机驾驶舱模型[31]，NASA的一个SPO模拟地面站[15]，上海交通大学的"驾驶舱＋地面站"SPO组合方案演示验证系统架构以及设计原型[9]。

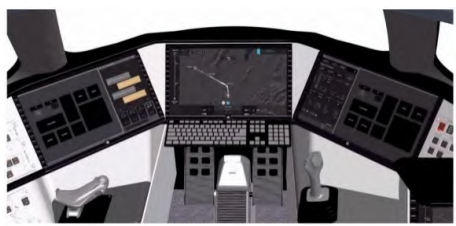

图1　Airbus的SPO飞机模拟驾驶舱[31]
Figure 1　An SPO flight deck simulator from Airbus
irbus'As deck minimum viable simulator

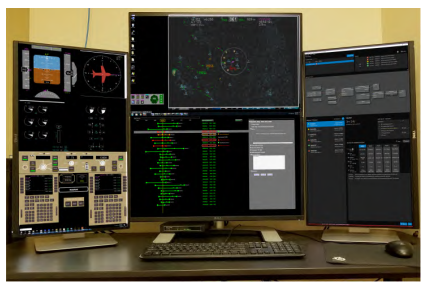

图2　NASA SPO实验模拟地面站[15]
Fig. 2　A NASA SPO simulated ground station

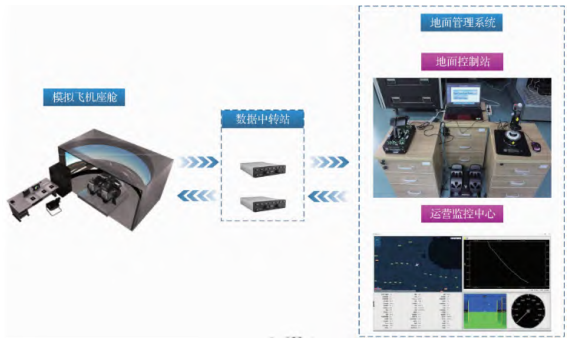

图3　上海交通大学的"驾驶舱+地面站"SPO组合方案演示和验证系统架构[9]

Figure 3　The demonstration and verification system architecture of the combined "flight deck+ground station" SPO solution proposed by Shanghai Jiaotong University[9]

## 1.2　研究方法

基于K.L.Vu,D.Schmid等人的综述[17,29],表2概括了目前SPO人因工程研究所采用的主要方法。人因工程研究方法是基于"以人为中心"的理念,在研究初期充分收集人(飞行员、地面站操作员等)的需求,然后构建设计概念和原型,通过实验获取用户(飞行员等)对设计概念和原型的反馈,整个过程是一个重复迭代的流程,最终筛选出符合人因工程需求的最佳SPO方案[13-14,24]。如表2所示,目前SPO人因工程研究总体上按照这样的"以人为中心"理念和方法展开,

表2　目前SPO研究中所采用的主要人因工程方法
Table 2　Major human factors methods used in the current SPO research

| 人因工程方法分类 | 文献数目 | 人因工程方法 | 方法结果描述 |
| --- | --- | --- | --- |
| 需求定义 | 16 | 研究综述,访谈,现场观察等 | SPO设计理念、系统需求、研发建议、研究策略等 |
| 概念定义 | 9 | SPO总体方案概念(初步),方案比较分析,可行性分析(初步)等 | 各种SPO研究方案、机载/地面站设备升级的初步设计概念等 |
| 人机功能和任务分析 | 27 | 认知工作分析(CWA),分层任务分析(HTA),工作领域分析(WDA),功能分析,流程分析,任务计算模型,操作事件序列图(OESD)等 | 在SPO操作环境中的不同飞行阶段,驾驶舱飞行员、地面站操作员以及设备系统的角色、功能、任务、流程等分析和定义 |
| 设计概念 | 6 | 低仿真系统原型化,低仿真人机界面原型化 | 认知飞机界面(CPAI),自适应认知飞行员-飞机界面(CPAI),剧本式人机界面(playbook interface),认知自适应人机界面(CAMMI)等 |
| 实验研究 | 13 | SPO模拟驾驶舱实验,SPO模拟地面站实验 | 在各种飞行场景中,利用现有双人驾驶舱或者低仿真SPO模拟驾驶舱、地面模拟站,比较不同SPO方案中飞行员(或者地面站操作员)的绩效,心理负荷等 |

这种"以人为中心"方法方法有助于优化人机匹配,支持适航认证,保证系统安全,降低后期系统开发的风险。同时,这样的方法也说明了在SPO系统研发中人因工程"预研先行"的重要性[13-14]。

## 1.3　阶段性研究结果

从研究的侧重面看,2012年的NASA SPO技术交流会确定了人因工程研究的5个重点领域:SPO设计方案、设备系统更新、人员交流与沟通、飞行员失能、适航认证[1]。目前的人因工程研究主要集中在前三个方面,取得了以下一些阶段性研究结果。

首先,基于有限的模拟驾驶舱实验研究,目前还没有足够的人因工程研究数据证实哪一种SPO方案具有明显的优势[17]。多项人因工程模拟舱实验研究表明,SPO驾驶舱机长借助于地面站或驾驶舱机载设备升级的支持,能够完成正常飞行场景和某些非正常场景中的飞行操作任务[17]。

第二,基于对飞行员工作负荷的考虑,没有地面站支持的SPO方案已经被许多研究放弃。多项模拟驾驶舱实验表明,在非正常或应急飞行场景中,飞行员普遍报告SPO会导致工作负荷增加,没有地面站支持的SPO方案差于有地面站支持的方案,地面站支持有助于降低飞行员工作负荷和飞行安全[14,18-19,25,29,32]。

第三,研究表明,地面站的多机"签派员"借助地面站设备,可以实现一人为多架正常起飞或进近的SPO飞机提供副驾驶职责的支持[17],这个结论是基于有效的地面站设备以及人机界面可以为

地面站操作员提供足够的有关 SPO 飞机的情景意识。

第四,从安全和人因工程等角度考虑,许多研究建议"驾驶舱方案+地面站方案"的组合方案[15-16,27,30],即通过提升现有驾驶舱机载设备,同时也为 SPO 飞行员提供地面站支持,构建一个"单一驾驶舱飞行员+驾驶舱机载设备+地面站操作员"协同实现的 SPO 新模式[9]。尽管该组合方案比较复杂,但研究者认为,这种组合方案可以处理各种不可预见的影响飞行安全的场景,最大程度地保证 SPO 飞行安全[27]。

最后,SPO 人因工程研究取得了一些阶段性结果,为 SPO 总体方案的初步选择提供了人因工程的学科支持[29]。Boeing 和 Airbus 等飞机制造商正在研发 SPO 民用飞机,由于缺少足够的信息,目前我们无法准确判断这些阶段性研究结果对制造商的 SPO 系统设计所产生的具体影响。

## 1.4　存在的主要问题

首先,尽管 SPO 人因工程研究取得了一些阶段性结果,但是目前的研究还不全面。虽然一些研究建议采用"驾驶舱方案+地面站方案"的 SPO 组合方案,但是该方案增加了系统总体设计的复杂性,涉及到人与机(驾驶舱,地面站)、空(驾驶舱)与地(地面站)人员之间功能和任务的重新分配,机载和地面设备的升级以及航空运输系统等多方面的重大更改,需要开展进一步的研究来为 SPO 系统总体设计提供完整的 SPO 人因工程解决方案。

其次,目前的 SPO 人因工程研究仅考虑了有限的飞行场景,还缺乏足够的研究证据来完整地验证 SPO 方案。例如,模拟舱实验仅考虑了有限的非正常和应急飞行场景,没有全面考虑如何从非正常飞行状态或者偏离正常自动化飞行程序的操作中恢复等复杂飞行场景[20],也没有充分考虑地面站"远程飞行员"为处于非正常飞行场景中的单架 SPO 飞机提供飞行支持的场景。这些高工作量场景对飞行员以及地面站操作员的工作负荷、人机交互和协作、飞行操控和决策等方面影响很大。另外,在 SPO "正常"巡航飞行阶段,低工作量可能会增加飞行员无聊和疲劳等效应,导致情景意识和警戒水平下降,目前还缺乏这方面的研究[27]。

最后,目前的 SPO 人因工程研究还没有开展(或者没有深入开展)针对一些 SPO 关键人因问题的研究。例如,飞行员失能[25],空地和人机之间协同操作和决策[33],适航认证等[6,29]。另外,许多实验采用不同的测试指标(工作负荷等),不利于跨研究的比较[34];研究大多采用现有驾驶舱设备或者低保真 SPO 驾驶舱设计原型,还没有具体的设计方案和高保真设计原型;研究缺乏对驾驶舱设备更新的总体考虑(提升现有机载自动化或者引进新的智能自主系统)[29,35-37];许多研究中的人机交互设计还停留在概念层面。

## 2　SPO 关键人因问题及研究重点

本部分针对一些 SPO 关键人因问题逐一进行详细综述和分析,找出目前研究中存在的不足之处,并且提出今后人因工程研究的重点。

### 2.1　人机功能和任务分配

作为人因工程研发的一项基础工作,人机功能和任务分配将为 SPO 的系统设计奠定基础。无论采用哪种 SPO 方案,都将对现有基于双乘员驾驶舱的人机功能和任务进行重新分配。SPO 研究首先需要优化人与机(机载设备系统)、空(驾驶舱)与地(地面站)人员之间功能和任务的重新分配。

目前人因工程研究已经开展了大量的这方面的研究[23]。如表2所示,研究者采用传统的功能分析、流程分析等方法,在不同飞行场景中,分析和定义 SPO 驾驶舱飞行员、地面站操作员以及设备系统的角色、功能、任务等;研究者还采用了一些新方法从不同的角度来分析 SPO 人机功能和任务的重新分配,例如认知工作分析,工作领域分析。但是,现有的 SPO 人机功能和任务分配研究工作还不完善,许多研究还侧重于方法的选择和比较,研究结果缺少人因工程实验验证,还无法为 SPO 方案的系统需求定义提供完整的分配方案[23]。

另外,目前的研究没有充分考虑地面站方案。例如,地面站方案的主要挑战之一是如何定义地面站操作员与 SPO 驾驶舱飞行员之间的功能和角色分配,但是大多数研究注重于地面站操作员的一种任务,例如支持高工作量情景下的单架飞机

或者多架正常飞行的飞机,只有 J. Lachter 等人的研究包括了地面支持的多任务组合[17,21]。

现有双乘员驾驶舱的人机交互设计主要是基于几十年前制定的人与自动化系统之间的功能分配方案,这种分配已被证明有时会阻碍飞行员履行其主要职责,不能有效地利用飞行员的潜能(处理异常状况等),浪费了飞行员的基本飞行技能[38-39]。这种设计是基于"以技术为中心"的理念,将一些飞行操作任务整体地分配给了自动化系统[40],造成了将飞行员置于"人在环外"的状况,容易降低情景意识,无法迅速有效地处理复杂的意外情况[38-39,41]。从方法论上来说,这种分配主要依据传统的 Fitts 模型[42]:人机功能分配取决于两者之间的相对优势,但是没有系统考虑飞行员潜能和技能[38],这个问题对 SPO 飞行员的影响更大。

今后针对 SPO 人机功能和任务分配的人因工程研究需要从以下几方面考虑。首先,吸取现有驾驶舱自动化系统设计的教训,采用"以人为中心"的理念来指导 SPO 人机功能和任务分配[13]。基于现有的研究,选择最佳的方法明确定义针对 SPO 的分配方案,要严格评估 SPO 中副驾驶工作量的分配去向[1]以及非正常和应急飞行场景中的分配方案,充分考虑工作负荷、人机交互决策模式、空-地协同的飞行操控和决策等因素,制定出完整的 SPO 人机功能和任务分配方案。

其次,采用人因工程实验(模拟舱实验等)验证 SPO 人机功能和任务分配方案,并且采用合适的指标(工作负荷等)来优化分配方案,为 SPO 方案的系统设计决策提供实验依据。

最后,在系统设计中考虑如何发挥飞行员潜能和技能。目前,在现有双乘员驾驶舱中,飞行员每次航线飞行平均花费不超过 10 分钟的手动操控[43]。SPO 驾驶舱自动化的提升会进一步减少手动操控,要考虑如何既能够发挥飞行员潜能和技能,又能保证飞行员"人在环内"的系统设计[38-39]。例如,采用基于自适应机制的智能系统,根据飞行员状态以及场景动态调整人机功能分配[38,44-45]。在低负荷操作中鼓励手动操控;在高负荷操作中,系统控制飞机,使飞行员能够执行航线规划或者应急任务[24]。

## 2.2 工作负荷

传统双乘员驾驶模式的一个重要特征就是工作分工,SPO 模式中将双乘员的工作由单一飞行员承担,增加了飞行员的工作负荷,这是 SPO 面临的重要问题之一。目前针对 SPO 工作负荷的人因工程研究主要集中在针对 SPO 驾驶舱飞行员和地面站操作员工作负荷的实验评估,以及如何通过设备系统的人机交互设计来降低工作负荷。

研究者对高工作负荷状态下如何保证 SPO 飞行安全的重要性已经达成了共识[1]。多项研究取得了基本一致的结果:在正常飞行场景中,多数飞行员可以接受 SPO 操作的工作负荷;而在非正常操作或应急飞行场景中,与双乘员驾驶舱相比较,飞行员普遍认为 SPO 的工作负荷明显增加[6][18-19][38][46]。SPO 方案需要给高工作负荷下的飞行员提供飞行支持方案(例如地面站支持)。

通过设备提升和人机交互设计来降低飞行员工作负荷是目前关注的一个问题。例如,认知飞行员-飞机界面(CPAI)设计概念[36],认知自适应人机界面(CAMMI)系统[47]。这些系统根据监测的飞行员工作负荷以及飞机状态,动态地调整人机任务分配(自动化水平等)。初步实验结果表明,系统可以降低飞行员的工作负荷。类似的人机自适应设计思路已经在早期研究中体现出来,例如驾驶舱支持系统(CASSY)[48-49]。其他的设计概念包括智能副驾驶电子交叉检查表[2],飞行员可穿戴技术[50]、操作告警系统[35]、增强现实眼镜[51]、虚拟飞行员辅助系统[52]、飞行操作推荐系统[53]等。

针对地面站操作员的工作负荷问题[21][22],N. Ho 等人的初步研究表明,在地面站智能设备系统(设计原型)的支持下,地面站签派员可以同时支持多架 SPO 飞机和管理地面高工作量的事件(机场关闭等)[54]。K. L. Vu 等人的研究表明当地面站"远程飞行员"为处于非正常飞行状态的单架 SPO 飞机提供支持时,就不应该执行其他职责[17]。

从长远来看,需要考虑未来空中交通管制(ATC)系统对 SPO 飞行员工作负荷的影响。例如,美国未来 ATC(NextGen)计划将一些 ATC 任务转移给飞行员,而且未来的空间将更加拥挤,这些变化有可能进一步增加 SPO 飞行员的工作负荷[1]。

目前针对 SPO 工作负荷的人因工程模拟驾驶

舱实验研究存在两个主要问题：其一，缺少对极端条件下飞行场景的考虑[20]；其二，实验研究采用的工作负荷测量指标不一致，例如生理、心理、脑电、眼动、主观量表等指标，给跨研究比较带来困难。

今后针对SPO工作负荷的人因工程研究需要考虑以下几方面工作。首先，建立SPO工作负荷评价体系，支持对SPO总体方案的实验研究筛选。基于适航认证要求（例如条款25.1523中的最小飞行组）[55]，选择合适的工作负荷测评指标，为SPO空地人-人与人-机交互、协同和决策的系统整体设计提供人因工程实验数据。

其次，针对目前研究中存在的问题，完善针对SPO工作负荷的人因工程模拟舱实验研究，特别需要评估在非正常和应急飞行场景中飞行员高工作负荷对SPO驾驶舱方案和地面站方案的影响，为SPO方案的决策提供实验依据，确保SPO驾驶舱飞行员代替现有双飞行员驾驶的工作负荷不超出安全范围。

最后，通过技术手段（自动化提升或者智能自主技术、人机交互设计）来降低飞行员工作负荷。例如，基于以往的研究，进一步开展智能机载人机功能自适应分配系统的研究，动态优化飞行员工作负荷。

## 2.3 飞行员失能

飞行员失能是指由于健康等原因飞行员丧失行为和认知等能力。传统双乘员驾驶模式下的飞行员失能可能危害飞行安全，飞行员失能对SPO飞行安全的影响更加突出。研究者对SPO飞行员失能的重要性已经达成了广泛的共识[1,16,56]，SPO驾驶舱必须装备机载感知系统来准确判断飞行员是否处于失能状态，一旦发现飞行员进入失能状态，机载自动化系统或者地面支持站必须快速接管SPO飞机。

发生飞行员失能的概率相当低，但是从单人驾驶特殊性、公众对SPO接受度以及民航安全等方面考虑，飞行员失能状态下保证SPO飞行安全显得尤为重要。有研究者建议，相对于双乘员驾驶舱双倍的余度式安全设计，SPO飞行需要有"三倍"的余度式安全设计[57]。

"未来欧洲环境中的飞机安全联盟"（SAFEE）研究项目针对飞行员失能和自杀事件的研究思路是开发飞机机载飞行员健康监测系统和飞机监控系统[58]。许多研究建议当SPO飞行员失能事件发生，地面站操作员应该接管飞机的操控，将飞机安全着陆在合适的机场，或者监控机载系统所操控的飞机自动着陆[16,21,25,59-60]。

目前针对SPO飞行员失能的人因工程研究还没有全面展开，这方面的文献有限，所关心的两个主要问题是飞行员失能监控指标和监控手段[1,25]。

D. Schmid的研究采用了系统理论事故模型和过程（STAMP）以及系统理论过程分析（STPA）[59]，研究分析了飞行员失能以及地面站如何检测这种场景来快速接管飞机操控。建模结果表明，使用升级的自动化系统可以防止由于SPO飞行员失能而可能导致的事故，该结果还没有经过实验验证。

在非SPO研究领域，人因工程针对人类操作员（飞行员、车辆驾驶员等）失能问题已经开展了许多研究[36]。从操作员失能的监控指标来说，人因工程研究建议采用对警觉、嗜睡、疲劳、失去知觉等状态的检测（心理、生理、眼动、脑电监测等）以及对高环境压力和工作量的检测（心理、生理监测等）等[16]。例如，Y. Lim等人[6]提出的SPO"虚拟飞行员助理"（VPA）系统架构建议采用大脑（例如血氧水平）、心血管（心率变异性等）、眼睛（眨眼率，瞳孔直径等）活动指标。

从SPO系统设计角度，研究者一般建议采用被动式和非侵入式飞行员监控方法[26]。有研究认为地面站的支持既能够帮助减轻SPO飞行员的工作负荷，同时在互动中可以帮助检测飞行员的状态[23]。

另外，当SPO飞行员失能事件发生时，SPO飞机实际上过渡到一架由地面站远程遥控的大型无人飞行器[6,23,26]。人因工程界已经开展了一系列针对大型无人飞行器应急状态下地面遥控的人机交互、人工接管等方面的研究，这些研究可以为SPO飞行员失能的系统解决方案提供支持[6,61]。

今后的人因工程研究主要包括以下三方面。首先，基于目前的研究，开展针对SPO飞行员失能监测指标的人因工程研究，利用人因工程实验研究来筛选最佳监测指标以及触发告警的最佳阈限值。

其次，开展针对SPO飞行员失能的机载监测

手段的人因工程研究。例如，监测系统的人机交互，系统舒服性等（非侵入式测量，远程监控等）[28]，监测准确性和预测性（脸部识别，脑电测量等）。利用针对自动驾驶车驾驶员监测的人因工程研究成果为SPO飞行员失能监控系统的开发提供参考。

最后，针对从飞行员失能事件发生到SPO飞机安全着落期间的平稳安全问题，人因工程要提供解决方案。解决方案包括飞行员失能监控、机载系统（自动化/智能自主系统）的自动接管、地面站紧急飞行支持、地面站操作员情景意识和角色转换等。

## 2.4 自动化与智能自主化系统更新

实现SPO的必要条件是更新现有设备系统，目前争议的焦点是采用自动化技术提升现有驾驶舱机载自动化系统还是引进基于智能新技术的机载智能自主化系统（autonomous systems）。

自动化系统依赖于固定的逻辑规则和算法来执行定义好的任务，当出现设计无法预料的飞行场景时，需要飞行员人工干预。大量的驾驶舱自动化人因工程研究发现，虽然自动化减少了飞行员的体力工作负荷，但是增加了飞行员自动化监控中的认知工作负荷，可能导致飞行员对自动化的过度信任、情景意识和警戒水平下降[62]。遇到意外事件时，系统可能引起飞行员的模式混淆、自动化情景意识下降等问题，这些问题导致了多起飞行事故[63-64]。

智能自主化系统通过基于人工智能算法、大数据和专家知识库（尤其是应对非正常和应急操作场景）的学习训练，在一些操作场景中，系统会具有一定的学习、自适应等能力，有可能在没有人工干预下独立执行一些设计中无法预期的场景任务，从而能在更大的操作范围内提供"自动化"功能[65]。不同于作为辅助工具的自动化系统，智能自主系统可以成为与人类合作的"队友"，分享任务和操控权，形成"人机组队"（human-machine teaming）式合作的新型人机关系，或者称为"人—自主化组队"（human-autonomy teaming）式合作[66-67]。

许多研究者希望在SPO驾驶舱引进一个"智能副驾驶"来承担起与SPO飞行员合作的队友角色，形成类似于双乘员驾驶舱的机组合作关系来解决SPO的一些挑战[1,18,37,45,68-72]。例如，Tokadli等人[37]采用一个"剧本委托界面"（playbook delegation interface，简称PDI）来评估SPO驾驶舱中的"人-自主化组队"式合作。该系统是一个基于领域知识库和决策-行为架构的智能自主系统，在一些设计无法预料的操作场景中可以辅助操作员。初步的实验结果表明，飞行员认为PDI有助于他们与该自主化系统的合作。Y. Lim等[6]提出的SPO"虚拟飞行员助理"（VPA）系统架构就是基于智能技术，该系统包括推理模型、不确定性分析模型、认知知识模型等，其设计目的是通过驾驶舱飞行员与智能自主化系统之间的协作来降低飞行员的工作负荷。

在NASA、FAA和Rockwell早期合作的一项SPO模拟舱实验研究报告中，针对SPO会导致飞行员工作负荷增加等问题，该报告建议SPO的技术干预方案不应该仅仅是提升现有驾驶舱的自动化系统，而是应该考虑引进新的智能自主化系统[18]。一些研究者也认同这样的技术路径[54]。

尽管目前针对"智能副驾驶"的技术和人因工程研究尚不成熟，但是在大型民用飞机机载设备领域已有一些正在研发的智能子系统，例如智能化推荐检查表及状态传感系统[71]，机载人机语音交互[73]，智能化空中交通防撞系统[2]，应对故障模式的智能飞行系统[2]，可穿戴智能设备[50]。这些研发有助于为SPO机载智能自主系统的技术发展路径提供支持。

SPO地面站同样需要设备系统更新。例如，NASA团队采用"人-自主化组队"式合作理念开展了一项针对地面站"签派员"执行飞行跟踪任务的评估[71]，该研究采用了一个自主约束飞行计划器（ACFP）系统。作为一种自动推荐系统，ACFP汇集多源信息，生成一份排序的选项（天气、位置、地形、飞机状况、机场跑道等），目的是通过该系统与地面站操作员的协作关系，支持地面站操作员在非正常场景中的快速决策。初步的模拟实验结果表明，与没有ACFP的地面站相比，参与者认为ACFP提供了足够的情景意识，降低了工作负荷。

提升现有的驾驶舱机载自动化系统来支持SPO是另一种思路。采用这种思路的SPO系统设计必须考虑双乘员驾驶舱自动化的人因工程问

题。如前所述,"以技术为中心"的理念导致了现有双乘员驾驶舱自动化系统容易引起飞行员"人在环外"的效应[38-39,65]。Bainbridge[74]发现了一个经典的"自动化讽刺"现象:自动化程度越高,操作员介入越少,对系统的关注度就越低;在应急场景中,操作员就越不容易通过人工干预来操控系统。更新 SPO 机载自动化必然提升系统的自动化程度,如何避免"自动化讽刺"现象是 SPO 系统设计的一个挑战。

针对驾驶舱自动化的升级方案,人因工程研究提出了自适应的设计概念,即根据飞行员状态,系统动态调整自动化水平[36,47]。其他已经研发或者正在研发的项目包括"电子副驾驶"[75]、"认知驾驶舱"(COGPIT)[76]、"机组人员驾驶舱自动化系统"(ALIAS)[77]、"数字化副驾驶"[78]等。尽管有些项目不是针对 SPO,但是这些设计概念为 SPO 驾驶舱自动化升级提供了参考。

人因工程研究表明,提升驾驶舱机载自动化的水平可以降低飞行员体力工作负荷,但是可能提升认知工作负荷并且降低情景意识。人因工程实验研究表明,这种情景意识损失可以通过使用中等程度的自动化来减少[79]。如何为 SPO 系统设计提供明确的解决方案有待于今后的人因工程研究。

针对 SPO 地面站自动化设备的升级方案,J. Lachter 等人[21]的研究评估了一个基于自动化技术的合作工具(CT)设计原型对 SPO 人机沟通和决策的影响。在该实验中,参与人员在三种配置中执行了非正常飞行场景:基准操作(双人飞行操作)、使用或者不使用协作工具的 SPO 操作。研究结果表明,虽然参与人员更喜欢基准操作的配置,但是他们认为协作工具的有助于 SPO 操作。这种从人机协作的角度来评估自动化升级值得鉴借。

综上所述,提升现有自动化系统还是引进智能自主系统,目前还没有明确的结论。今后的人因工程研究需要解决以下几方面问题。

首先,无论是采用自动化或者智能自主系统,我们强调采纳"以人为中心"的理念来指导对 SPO 设备系统的总体设计。该理念要求将人类操作员放在系统研发的中心位置考虑,发挥人类智能与机器智能间的优势互补,实现"人在环"的系统设计,保证人类操作员拥有对 SPO 飞机的最终操控权。

第二,根据人因工程研究、机载自动化技术和智能自主技术可行性,我们初步建议 SPO 驾驶舱设备系统采用"自动化 + 自主化"的组合式方案,根据场景复杂性选用技术,利用两种技术的优势互补来获取最大的安全保证。例如,提升现有机载自动化来开发面向一般飞行环境的自动飞行模式,引进智能自主系统来开发面向复杂飞行环境的自主飞行模式(可独立执行一些无法预期的任务)。人因工程要从人机功能分配、工作负荷、人机交互协同等方面出发,优化人-自动化-智能自主系统三者之间的整合设计,通过实验验证最终的技术方案。

第三,针对驾驶舱设备升级,开展人机交互、协作和决策的人因工程研究,为 SPO 空地人-人与人-机交互、协同和决策的系统整体设计提供人因工程方案。例如,基于智能技术,构建和实验验证基于"人-自主化组队"合作的人-自主化之间的人机交互和决策模式及设计概念;在以往人因工程研究基础上,构建和实验验证 SPO 驾驶舱飞行员与自动化之间的人机交互决策模式和设计概念。

最后,针对驾驶舱机载自动化升级设计,系统设计需要避免"自动化讽刺"现象,解决目前双乘员驾驶舱中的人因问题(简化自动驾驶模式,避免"人在环外"等),在人与自动化(自动化水平、人机功能分配、工作负荷等)之间找到一个最佳设计平衡点,保证只有通过严格人因工程实验验证的自动化升级方案才能在 SPO 方案中被考虑。

## 2.5　人机交互和人机界面

驾驶舱和地面站设备系统要在 SPO 中发挥作用离不开有效的人机交互。回顾历史,目前双乘员驾驶舱人机交互设计基本上遵循"以技术为中心"的理念,导致人机界面拥有过分复杂的自动化模式和控制方式,带来产生人为差错的隐患[80-81]。例如,垂直导航(VNAV)操作中的众多自动化控制方式(垂直速度-V/S,飞行高度改变/FLCH, VNAV 航路,VNAV 速度,飞行路径角/FPA 等)增加了飞行员的认知工作负荷;新增机载设备的告警信号(TCAS,EGPWS 等)没有与原有机载告警信号实现有效整合,容易给高负荷状态下的飞行员造成信息过载[81]。

受跨机型设计通用性和兼容性、飞行员培训、适航风险等因素的影响，后续新机型的驾驶舱人机交互设计并没有得到根本的改进。SPO应该是优化驾驶舱人机交互设计的一个新机遇，必须考虑如何解决这些"历史遗留"的人机界面设计问题。

现有双乘员驾驶舱显示过多与飞行员当前任务无关的信息，这种信息过载问题对SPO飞行员的影响将更大。现有驾驶舱人机显示界面实际上并没有完全实现从"空分制"（空间固定式机电仪表）向"时分制"（实时动态化数字显示器）显示方式的过渡，并且缺乏基于优化等级的动态信息显示。Airbus考虑在SPO驾驶舱中采用基于"以飞行员为中心"的动态优化显示方式[31]，即围绕飞行员任务的需求，显示关键信息，降低次要信息的优先等级。Airbus认为SPO机载人机界面要保证飞行员拥有合适的工作负荷水平，例如在巡航低负荷阶段，人机界面需要为飞行员提供一定的人机交互活动，保持"人在环"的状态；在高负荷阶段，人机界面要突出当前飞行目标参数的显示[31]。在模拟器上针对飞行员眼动扫描行为的实验结果支持Airbus的思路[49]，该研究表明SPO操作会增加飞行员的视觉工作负荷。

目前针对SPO驾驶舱人机交互的人因工程研究有多种思路。如前所述，基于"自适应设计"理念的认知飞行员—飞机界面（CPAI）[36]、认知自适应人机界面（CAMMI）[47]；基于"人—自主化组队"式合作理念的"剧本委托界面"（PDI）[37]；基于人机交互技术的可穿戴技术和语音交互[50,73]。

针对SPO地面站设备的人机交互设计，有研究认为地面站设备的人机交互设计需要采用SPO机载人机界面的镜像显示方式[25]，即显示与SPO驾驶舱一致的信息，这有利于地面站人员通过设备的人机界面来远程操控SPO飞机。Y. Lim等提出的虚拟飞行员助理系统（VPA）系统架构就包括了SPO飞行员监控、飞行管理/控制等功能的人机界面[6]。另外，NASA团队采用了一个"应急着陆协调"系统（ELP）[53,82]，当SPO飞机发生故障时，作为推荐系统的ELP可帮助飞行员选择最佳紧急着陆机场。实验结果表明，地面站操作员认为该推荐系统人机界面的透明化设计能够为决策提供更好的解释，并且能有效支持在非正常场景中高负荷的地面站操作。

如何通过有效的地面站设备人机交互设计来帮助地面站操作员保持足够的情景意识也是人因工程关心的一个问题。S. L. Brandt等的初步研究表明，既使地面站操作员缺乏对某一SPO飞机飞行状态的详细了解，但是当该机向地面站请求副驾驶职责的飞行支持时，地面站操作员借助地面站设备人机界面仍然可以迅速获取足够的情景意识来为该机提供支持[20]。J. Lachter等的研究表明[83]，地面站设备的透明化人机界面可以及时将SPO飞行环境和系统数据提供给地面站操作员，从而获取足够的情景意识。有研究建议利用可穿戴人机交互技术可以帮助SPO飞行员提高情景意识[50]。

总的来说，目前针对SPO设备的人机交互研究目前还不完整，SPO空对地人机协作和决策系统涉及到空地人—人和人—机之间的交互、协同和决策，人—自动化/自主化交互等一系列人机交互问题，有待于今后进一步的人因工程研究。

一方面，SPO为新型驾驶舱人机交互的优化设计提供了一个新的机遇，SPO驾驶舱人机交互设计要基于"以飞行员为中心"的理念，解决现有双乘员驾驶舱人机交互存在的问题。例如，基于飞行员任务的人机界面动态优化显示方式，简化驾驶舱自动化控制和显示方式。要考虑采用创新方法，例如基于"人—自主化组队"式合作的人机交互设计。

另一方面，人机交互和界面设计要考虑多种因素（工作负荷，飞行员失能等），SPO研发可以构建多个设计原型方案，通过人因工程模拟驾驶舱实验来优化人机界面的设计。

## 2.6　人—机与空—地的飞行协同操控和决策

SPO改变了驾驶舱飞行员的认知决策模式。一方面，SPO避免了现有双乘员驾驶舱中双人飞行员之间潜在的认知决策冲突，有助于提升决策效率；另一方面，SPO驾驶舱飞行员将更多地依赖于个人知识、人-机（机载系统）之间以及人-人（地面站）空地之间的协同操控和决策，因此SPO的飞行操控和决策模式可能变得更为复杂。针对这种新型的飞行操控和决策模式，人因工程研究还没

有通过深入的研究来达到共识。要达到这样的共知,人因工程研究首先要回答三个关键问题。

第一个问题是关于SPO驾驶舱飞行员的角色转变。民用飞机飞行员的主要任务是飞行、导航、通信(A-N-C),驾驶舱自动化的引进增加了一个管理系统的任务,并且带来了一些不可预见的事件和故障模式,需要飞行员处理,因此,飞行员更像是一名"自动化管理员"或者"异常处理员"[28]。S. M. Sprengart等人认为,未来的机载设备系统将自行管理具体的飞行任务,飞行员可能转变为飞行管理角色(从机场A的机场B的任务等),成为一名"任务经理"。Airbus也认为未来驾驶舱中的飞行员一般不用从事具体的航线飞行操作任务,而主要将承担航线规划等"任务管理"的角色[30]。

SPO涉及到在驾驶舱中移除和替代第二名飞行员的核心问题,绕不开一个问题:如何定义人类操作员的角色。SPO驾驶舱飞行员的这种角色转变和定位将影响人机和空地之间的功能分配、协同操控和决策模式,影响SPO的系统技术方案。

第二个问题是关于SPO飞机的最终飞行操控决策权。基于"以技术为中心"理念的机载自动化升级"促使"飞行员成为系统管理员,部分地导致目前双乘员驾驶舱中飞行员与自动化之间在飞行操控和决策方面存在的一些不匹配,导致了与自动化相关的人因问题[13,41]。

许多SPO研究出于对SPO整体设计复杂性的担忧,研究的重点只是在一个"舒适区"内针对SPO方案提出一些针对现有系统的局部改进,这是一种"演变"而不是"革新"的设计路径。虽然这种"演变"路径对系统开发和适航认证成本要低得多,但是很难从根本上解决目前人-自动化交互中的问题,而且长远来看,可能带来飞行安全、人因工程、系统升级可扩展性等方面的一系列问题。

S. M. Sprengart等认为,离开"舒适区"是寻找SPO系统方案的必要条件,只有这样才能将SPO飞行员与机载系统(自动化、智能自主化、人机交互等)在人机与空地的飞行协同操控和决策等方面达到最佳的匹配。无论哪一种SPO方案都需要适航认证,SPO提供了一个独特机遇。S. Neis等建议[27],SPO系统方案的设计应该回到原点,将人类操作员放回中心,让技术适应人类操作员,最终达到人机交互、协同操作和决策的最佳匹配。

第三个问题是关于飞行的操控决策权管理和权限分配。现有双乘员驾驶舱的设计是基于"决策控制权在机长手中,直到移交给另一人为止"的原则[33]。SPO飞行操控权的授权管理和权限分配可发生在人-机(机载、地面站系统)之间或者人-人(驾驶舱与地面站)之间,这个过程可能会出现飞行操控决策和权限分配方面的冲突[27]。目前的研究还没有明确地定义相应的分配原则和模式。

有研究建议只要SPO系统设计定义了相应的分配原则和模式,我们可以利用智能自主技术来实现对SPO飞行操控权限的分配。例如,智能化空中交通防撞系统(ITCAS)在检测到即将发生碰撞时,并且飞行员失能或无法及时做出反应时,系统可以自适应调整拥有的权限级别来接管飞行操控[2]。

人机与空地的飞行协同操控和决策模式设计还需要考虑一些人因问题。例如,人因工程研究已经表明,当飞行员处于环外状态时,其控制权应该被收回[63];当飞行员从系统中收回飞行操控权时,系统设计需要考虑飞行员认知延迟的影响,尤其在应急状况中对系统的诊断会出现更长的认知延迟[84]。研究表明,在应急场景中地面站操作员利用人机协作系统可以有效支持空地之间操控权的分配[76]。

综上所述,目前还没有足够的SPO人因工程研究能够完整地回答这些重要问题。今后的研究首先需要构建和实验验证SPO机长、驾驶舱设备系统和地面站操作员三方空地协同的飞行协同操控和决策模式。这种模式首先要明确以上三个问题的答案,考虑各种飞行条件以及SPO飞行员能力(表1),考虑在操控权分享和转移过程中潜在的人-人或者人-机冲突[1],并且通过人工程研究来验证。

另外,今后的研究还需要从SPO人机功能的重新分配出发,严格定义飞行操控和决策权限的优先等级分配方案[33]。例如,如果机长离开驾驶舱,地面站承担"副驾驶"的支持人员是否可以履行机长的职责;在飞行员失能情况下,飞行权限是否自动转移到地面"副驾驶"或者由机载系统接管;如果需要地面站远程控制,地面站操作员如何知道他们必须接管控制权?谁拥有最终操控权;作为一种罕见的情况,如果同时发生地空两名飞行员以敌对或自杀的方式行事,是否需要考虑备

份地面站操作员[28]。

## 2.7 人—人交流与沟通

SPO的系统方案需要考虑地面站操作员、ATC管制员和驾驶舱的一体化协同模式[9]，在这一模式中，人—人之间的有效交流与沟通至关重要[1]。

SPO驾驶舱飞行员与地面站操作员之间的交流除了技术挑战以外，不同空间上人-人（SPO机长与地面站操作员）之间的沟通、协调和决策是一个重要的人因问题，事关飞行安全。由于无法获得非语言线索和动作的信息，飞行员的空间分离可能会对相互之间的交流以及情景意识产生负面影响，例如，飞行员对他们的角色（谁在飞行？）以及动作是否完成（是否输入了命令？）可能会感到困惑。

J. Lachter等的实验结果表明，在一些非正常飞行场景中，虽然SPO飞行员不喜欢空间分离的操作，但是在空间分离与不分离两种条件下的SPO飞行绩效没有显著差别[76]，因此，研究者认为空间分离对SPO空地合作的影响没有比预料的大。该模拟舱实验还比较了SPO机长与地面站"副驾驶"的协同操作（"空间分离"方案）、现有双人驾驶舱操作（"无空间分离"方案）、有或没有机组资源管理（CRM）协作工具支持的三种实验条件，结果表明，在一些非正常飞行场景中SPO驾驶舱飞行员都能够完成飞行任务，但是他们更倾向于采用"无空间分离"的方案以及有CRM协作工具的支持[21,83]。

缺乏副驾驶的SPO驾驶舱由于缺乏舱内人-人之间的交流与沟通，飞行员容易产生单调、低警戒、易疲劳等状态，因此目前的研究都建议SPO应该被限制在短途航线运营中[1]。现有的SPO人因工程研究还没有充分考虑SPO对飞行员疲劳的影响以及针对飞行员疲劳管理的解决方案[85]。Airbus认为SPO机载人机界面设计应该考虑飞行员的疲劳管理[31]，例如在巡航低负荷阶段，人机界面为飞行员提供一定的人机交互活动，保持一定的警觉水平。

今后的人因工程研究需要考虑以下三个方面。首先，进一步评估在不同SPO飞行阶段中空间分离对飞行员的工作负荷、情景意识、警戒水平以及单调状态等方面的影响，为SPO人机与空地的协同操控和决策模式的系统设计提供人因工程的支持。

其次，在系统设计中，通过技术和人机交互设计等手段来降低SPO空地之间潜在的冲突风险，通过有效的、透明化的人机交互界面来促进人-人以及人机之间的沟通[1]。

最后，提供针对冲突管理的人因工程解决方案。例如，制定冲突管理的策略和方法[1]，改进现有CRM方法，修改现有飞行程序。

## 2.8 人为因素适航认证

SPO的适航认证是一个公认的棘手问题[6,23]，其中，美国联邦航空总署（FAA）和欧盟航空安全局（EASA）的态度尤为重要。在2012年的NASA SPO会议上[1]，FAA飞机与飞行机组界面部主任Steve Boyd表明：SPO研发不应该将适航认证视为一个障碍，目前的适航认证是达到飞行安全目的的一个不完美的工具；如果新技术出现，在保证安全的前提下，FAA会认为某些条款是过时的。EASA目前正在考虑放宽对大型民用飞机SPO适航认证的限制[86]。

适航条款要求本身是与时俱进的。例如，80年代继Boeing 757采用双人操作驾驶舱以后，Boeing成功说服FAA将原先计划的Boeing 767的3人制驾驶舱升级为双人制驾驶舱。另外，适航条款FAR 25.1523发布在上世纪40年代[87]，该条款的适航要求是基于飞行绩效，并没有明确指定最小飞行机组人数。在60年代，该条款增加补充了飞行操作、导航、通讯等6方面的工作负荷要求。

从人为因素（人因工程）适航认证角度看，FAR-25部中的某些适航条款可能是SPO适航认证的潜在障碍[55]。例如，FAR 25.1523（最小飞行机组）包含与飞行员失能相关的要求，并将事故的原因归因于单人驾驶操作，这可能表明适航当局不愿对SPO进行认证。虽然25部没有明确说明SPO是不可以适航认证的，但是25部中所用的语言是假定双人操作的驾驶舱[1]。有研究者质疑，虽然所有的适航规范都规定大型民用飞机驾驶舱必须有不少于两人的机组，但是同时又规定所有飞机必须可以由一名飞行员从任一座位上操作，这是否表明现有双乘员驾驶舱已经满足了对SPO的适航认证要求[24]。

针对人为因素适航条款，人因工程已经开展了一些研究。例如，FAR 25部中许多适航条款从人因工程角度对驾驶舱设计提出了保证最低安全

的适航要求[87],为进一步降低由人为因素导致的事故发生率,国际航空界一直在努力将航空产品的人因工程设计高于适航要求,并且形成了以下两方面的共识。

一方面,目前的人为因素适航条款内容滞后于机载技术和人因工程领域的发展。例如,许多条款内容是基于以往机电式机载设备而制定,没有及时反映当代数字化、自动化和综合化机载人机交互技术对人因工程设计提出的更高要求,由此带来条款内容多侧重于物理空间等方面的基本要求(可达性、可视性等)。人为因素条款基本按照"以系统为中心"的方式分散地罗列在相关的系统和部件条款中。

另一方面,人为因素适航条款缺乏对驾驶舱人机交互设计的指导作用[88]。例如,自动化驾驶舱可能引起飞行员的自动化方式混淆和选择的人为差错[39,64]。尽管设计不能(也不可能)完全消除人为差错,但作为保障最低飞行安全的适航标准,如何从设计上帮助最大限度地减少人为差错有待进一步完善。另外,人为因素适航条款缺乏对不同飞机制造商之间的驾驶舱人机界面基本元素的一致性设计指导,目前的双乘员驾驶舱存在不一致的界面设计(显示画面格式、起飞/复飞开关位置、自动油门断开装置、自动飞行方式板布局和方式命名等)。

更重要的是,从进一步提高飞行安全的角度看,目前的人为因素适航条款缺乏对新一代驾驶舱人机界面优化更新的前瞻性指导[88]。这个问题对SPO研发影响更大,因此适航认证条款应该与时俱进,从而能够有效地指导SPO的人因工程设计。

满足当前的适航条款对SPO是一项挑战。例如,就条款25.1302(人误管理)的适航要求来说,在多人驾驶舱中,事故数据表明一名飞行员所产生的人为差错经常由另一名飞行员发现和改正,SPO则失去了这种机会,这是否会给安全造成影响[1,38]?根据条款25.1523,要证明SPO中单飞行员的工作量(借助某种支持方案)等于或小于现有驾驶舱双人飞行员的工作量也是一个挑战。

人因工程研究表明,建立起申请人与局方(适航当局)之间有效的协调关系非常重要[88]。局方应主动地参与型号研制,指导申请人的人为因素前期认证工作;而申请人要主动争取局方指导。

SPO研发可能带来驾驶舱人机界面的重新设计,尽早建立起申请人与局方的协调合作关系尤其重要。

另外,SPO对飞行员如何获得飞行资质和适航认证也带来了新问题。目前,一名飞行员要获得双乘员驾驶舱的飞行资格,首先要成为一名观察员,积累一定的飞行小时数后才能成为一名副驾驶,再积累一定的飞行小时数后才能成为一名机长。若引入SPO运营机制,飞行员的带飞培训、驾驶舱飞行操作资质获取等方面的工作将面临新的问题。

尽管SPO人为因素适航认证是人因工程界非常关心的一个问题,但是还没有进入实质性的研究阶段,下一步人因工程工作可以从以下几方面考虑。

首先,将SPO的人为因素适航意识贯穿在人因工程的研发全流程中。在研究初期,确定所有适用的人为因素条款,将各条款的具体要求分解为各系统和部件的设计指标,对不符合现有条款要求的设计做到早发现早沟通。同时,SPO研发要挑战过时的现有条款,尽早获取申请人与局方之间的共识。

其次,SPO人因工程研究要与适航条款挂钩,实验指标的选取尽量与适航取证要求相符合,有利于获取有效的验证数据,提高适航认证的效率。例如工作负荷、飞行员失能等方面的实验研究。

最后,将SPO人为因素适航认证纳入机载设备供应商的选择决策以及合作中。作为系统整合的飞机制造商,型号认证是对驾驶舱系统整体设计的认证[88],供应商的单一机载部件设备获取适航认证并不能保证对驾驶舱设计的整体认证,因此人因工程在设备供应商选择和合作中要有话语权是保证驾驶舱系统整体设计获取适航认证的手段之一。

## 3　今后SPO人因工程研究思路和建议

综上所述,目前SPO人因工程研究取得了一些阶段性研究结果,但是还有许多SPO关键人因问题还没有开展或者需要进一步的深入研究。针对今后人因工程研究,我们提出以下总的思路和建议。

## 3.1 SPO的设计理念

今后SPO的研发工作需要继续强调"以人为中心"的理念,同时,我们具体定义了"以人为中心"的SPO设计理念和人因工程设计指导原则(见表3)。

表3 "以人为中心"的SPO设计理念和人因工程设计指导原则(部分)
Table 3 The "human-centered" design philosophy for SPO and the guiding principles of human factors engineering (examples)

| "以人为中心"的SPO设计理念 | SPO的人因工程设计指导原则(部分) |
| --- | --- |
| 充分考虑人的需要、能力、潜力和极限,利用人已有的经验、知识和技能 | · 人机功能和任务的重新分配应该遵循"以人为中心"的理念以及基于人机协同合作的方法,而不是"以技术为中心"的理念<br>· 基于自动化或智能自主化技术的机载系统不应该取代飞行员,应该有针对性地发挥和增强飞行员的潜能(处理异常情况等)以及技能(基本飞行技能等)<br>· 地面站操作员的职责升级设计(签派员、港口飞行员、远程飞行员)要充分考虑人员的经验、知识和技能<br>· 最大限度地降低SPO驾驶舱飞行员在非正常和应急飞行场景中的工作负荷 |
| 人拥有对SPO飞机的最终操控权 | · 人(驾驶舱飞行员或地面站操作员)拥有对SPO飞机的最终操控权<br>· 除非发生意外状况(飞行员失能等),SPO驾驶舱飞行员拥有对飞机的最终操控权<br>· 系统设计应保证应急状态下的有效的余度化操控权限设计。例如,如果飞行员失能发生,机载(自动或智能自主化)系统及时找到合适机场,启动应急着陆系统,控操飞机自动着陆,地面站操作员全程实时监控。若综合相关信息和充分证据表明,地面站操作员的介入是必要的,地面站操作员具有最终操控权,可随时接管SPO飞机,并且控操飞机安全着陆 |
| 技术既是支持人作业的工具,也是与人合作的团队队友 | · 驾驶舱机载智能自主化系统不仅是支持飞行员的一个辅助工具,而且也是一名与飞行员合作的团队成员(例如,采用"智能副驾驶"系统来承担人类副驾驶的一些职责)<br>· 人机之间可以分享情景意识、任务、目标和飞行操控权等<br>· 吸取驾驶舱自动化设计的教训,保持飞行员有效监控和在环状态,保证当应急状态发生时飞行员能够快速有效地接管飞行操控权 |

SPO设计理念对于SPO研发以及飞行安全极为重要,它将贯穿于SPO系统研发的整个生命周期。表3所定义的"以人为中心"SPO设计理念将有利于各专业人员统一设计思路,将对人机(自动化和智能自主系统)功能分配、工作负荷、人机与空地的飞行协同操控和决策、人机交互决策模式、人机交互设计(例如简化、自适应化、余度化和容错化)、SPO飞机最终操控权等方面的系统设计和决策起着重要的指导意义,避免以往驾驶舱机载系统研发中由于采用"以技术为中心"理念所带来的问题。Boeing和Airbus在开发新机型产品的初期都首先制定了"以人(飞行员)为中心的中心"驾驶舱设计理念[13-14]。另外,表3中的SPO设计理念和设计指导原则将在今后的工作中将不断完善。

## 3.2 今后SPO人因工程研究总的思路

根据前面的综述和分析,围绕构建一个"单一驾驶舱飞行员+驾驶舱机载技术系统+地面站操作员"协同实现的SPO新模式,本文从人因工程角度出发,对下一阶段的SPO人因工程研发提出以下一些初步建议。

首先,我们强调"以人为中心"的理念来指导SPO研发,制定具体的人因工程设计原则来指导研发工作,并且将人因工程方法整合在SPO研发的整个流程中[12,51]。同时,研发人员要将适航认证贯穿于SPO的研发中,要特别注重一些重要适航条款的要求,例如25.1523(最小飞行机组)、25.1302(人误管理)以及与飞行员失能等相关的条款。在现有研究的基础上,优化人因工程实验测试指标(例如工作负荷指标),为今后适航认证做准备。

其次,围绕构建一个"SPO驾驶舱飞行员+地面站操作员+机载和地面站设备支持系统"协同实现的SPO新模式,深入开展针对"驾驶舱方案+地面站方案"SPO组合方案的人因工程研究。研究重点集中在更大飞行操作环境范围内(非正常和应急飞行场景中)验证SPO方案的安全性。开展针对工作负荷、飞行员失能、人机与空地的飞行协同操控和决策、人与自动化/智能自主系统交互、协同以及决策支持等关键问题的人因工程研究,为SPO系统整体设计提供人因工程的解决方案。

第三，构建SPO人因工程实验测试平台。该实验测试平台应该主要包括模拟驾驶舱(模拟各种非正常和应急飞行场景的主要航线飞行任务)，模拟地面站(模拟对SPO飞机提供主要的支持任务)，人因工程测试设备(例如眼动仪，脑电、生理测量仪)以及数据采集储存系统等。

最后，开展跨学科、跨行业的协同合作。SPO研发需要企业与科研院校、企业与设备供应等各方面的人因工程合作研发。针对SPO驾驶舱与地面站的组合研究方案，还需要与航空公司、民航空管机构等单位协同合作。

# 4 结束语

针对SPO的人因工程预备研究已经展开，最初的研究主要集中在驾驶舱机载设备升级和地面站支持两种SPO方案。基于有限的人因工程模拟驾驶舱实验，目前还没有足够的数据证实哪一种SPO方案具有明显的优势，同时也没有发现可以完全阻止实现SPO的障碍。但是，没有地面站支持的SPO方案已经被许多研究放弃，许多研究建议采用"驾驶舱方案 + 地面站方案"的第三种SPO组合方案。

尽管SPO人因工程研究取得了一些阶段性结果，但是目前的研究还不全面，需要开展进一步的研究来为SPO系统总体设计提供完整的SPO人因工程解决方案。针对一些SPO人因工程关键问题，本文逐一进行详细的文献综述，分析目前研究中存在的不足之处，提出今后研究的重点。最后，对下一步的SPO研发工作提出了一些初步建议。

作者简介:

许 为(1958— ),男,博士,教授。主要研究方向:人—人工智能交互、航空人因工程、人机交互等。

陈 勇(1967— ),男,研究员,总设计师。主要研究方向:飞机总体气动、结构、安全性等。

董文俊(1983— ),男,博士,高级工程师。主要研究方向:民用飞机驾驶舱设计、人为因素评估与仿真等。

董大勇(1976— ),男,博士,研究员。主要研究方向:民用飞机驾驶舱设计,民用飞机驾驶人为因素设计与评估等。

葛列众(1956— ),男,博士,教授。主要研究方向:人因工程、人机交互、面部认知等。

(编辑:马文静)